\documentclass[fleqn]{2023SCGE}
\usepackage{tabularx}
\usepackage{longtable}
\usepackage{extarrows}
\usepackage{supertabular}
\usepackage{threeparttablex}
\usepackage{graphicx}
\usepackage{dcolumn}
\usepackage{bm}
\usepackage{amsmath}
\usepackage{overpic}
\usepackage{booktabs}%
\usepackage{makecell}
\usepackage{enumerate}
\usepackage{multirow}
\usepackage{rotating}[figuresright]
\usepackage{booktabs}
\usepackage{lineno}
\usepackage{hyperref}

\newcommand{\mnras}{\textsc{Monthly Notices of the Royal Astronomical Society}}
\newcommand{\apj}{\textsc{The Astrophysical Journal}}
\newcommand{\apjl}{\textsc{The Astrophysical Journal Letters}}
\newcommand{\nat}{\textsc{Nature}}
\newcommand{\aap}{\textsc{Astronomy \& Astrophysics}}
\newcommand{\apss}{\textsc{Astrophysics and Space Science}}
\newcommand{\pasa}{\textsc{Publications of the Astronomical Society of Australia}}

\newcommand{\araa}{\textsc{Annual Review of Astronomy and Astrophysics Annual Reviews}}

\newcommand{\FIG}[1]{Fig.~\ref{fig:#1}}
\newcommand{\TAB}[1]{Table~\ref{tab:#1}}
\newcommand{\SEC}[1]{Section~\ref{sec:#1}}

\begin{document}

\ensubject{subject}

\ArticleType{Article}
\SpecialTopic{SPECIAL TOPIC: }
\Year{????}
\Month{??}
\Vol{??}
\No{?}
\DOI{??}
\ArtNo{000000}
\ReceiveDate{?? ??, ????}
\AcceptDate{?? ??, ????}

\title{Detection of Cyclotron Absorption in the Radio Emission of GPM 1839-10}

\author[1,2]{Yunpeng Men}{{ypmen01@gmail.com}}%
\author[1]{Ewan Barr}{}
\author[3,4]{Yuanhong Qu}{}
\author[5]{Csanád Horváth}{}
\author[1]{\\Jinchen Jiang}{}
\author[1]{Gregory Desvignes}{}
\author[5]{Natasha Hurley-Walker}{}
\author[1]{Michael Kramer}{}
\author[6]{\\Rui Luo}{}
\author[5]{Samuel J. McSweeney}{}
\author[1]{Jason Wu}{}

\AuthorMark{Yunpeng Men}

\AuthorCitation{Y. P. Men, et al}

\address[1]{Max-Planck-Institut f\"ur Radioastronomie, Auf dem H\"ugel 69, D-53121 Bonn, Germany}
\address[2]{Department of Astronomy, School of Physics, Huazhong University of Science and Technology, Wuhan, 430074. China}
\address[3]{Nevada Center for Astrophysics, University of Nevada, Las Vegas, NV 89154, USA}
\address[4]{Department of Physics and Astronomy, University of Nevada Las Vegas, Las Vegas, NV 89154, USA}
\address[5]{International Centre for Radio Astronomy Research, Curtin University, 1 Turner Ave, Bentley, WA, 6102, Australia}
\address[6]{Department of Astronomy, School of Physics and Materials Science, Guangzhou University, Guangzhou 510006, China}


\abstract{GPM\,1839$-$10 is an intriguing long-period radio transient (LPT), distinguished by its activity spanning at least three decades and its highly unusual emission characteristics. These features include orthogonal polarization mode (OPM) switches, down-drifting sub-structures, and distinct linear-to-circular polarization conversion behaviors. In this work, we present follow-up observations utilizing the FAST telescope at L-band, yielding a total of seven detected radio pulses. We find a consistent association between OPM switches and a decrease in polarized intensity. This feature strongly supports the hypothesis that the OPM switches are generated by the incoherent summation of OPMs. Our measured Rotation Measures (RMs) are consistent with previous observations, indicating that the magneto-ionic environment is stable. If the source is in a binary system, such stability suggests it may host a weakly magnetized companion. Crucially, we firstly observe clear evidence of a cyclotron absorption feature in one radio pulse, a signature rarely observed in radio sources. This feature allows us to infer that the magnetic field strength at the absorption site has a lower limit of tens of Gauss, which is necessary for the phenomenon to occur. This characteristic can be explained in a scenario where GPM\,1839$-$10 possesses a  weakly magnetized companion star.}


\PACS{???, ???, ???}

\maketitle

\begin{multicols}{2}
\section{Introduction}\label{sec:introduction}
Long-period Radio Transients (LPTs) are a newly identified class of objects that emit sporadic radio pulses with periods exceeding one minute. To date, a total of 15 LPTs have been reported, exhibiting periods ranging from 76 seconds to 6.45 hours~\cite{Anumarlapudi2025MN, Dong2025ApJLb, Dong2025ApJLa, McSweeney2025MN, Bloot2025AA, Wang2025Nat, Ruiter2025NatAs, Lee2025NatAs, Hurley-Walker2024ApJL, Dobie2024MN, Li2024arXiv, Caleb2024NatAs, Pelisoli2023NatAs, Hurley-Walker2023Nat, Caleb2022NatAs, Hurley-Walker2022Nat, Marsh2016Nat, Hyman2005Nat}.
\Authorfootnote
Some of these LPTs have been identified as white dwarf (WD)–M dwarf binary systems, such as ILT\,J1101+5521~\cite{Ruiter2025NatAs}, GLEAM-X\,J0704$-$37~\cite{Hurley-Walker2024ApJL}, J1912$-$4410~\cite{Pelisoli2023NatAs}, and AR\,Sco~\cite{Marsh2016Nat}.
In contrast, PSR\,J0901$-$4046 and CHIME\,J0630+25 are likely highly magnetized neutron stars, with surface magnetic field strengths of approximately $10^{14}$\,G, inferred from their measured periods and period derivatives under the assumption of magnetic dipole braking~\cite{Dong2025ApJLb, Caleb2022NatAs}. In addition to radio emission, several LPTs also exhibit X-ray counterparts, such as ASKAP\,J1448$-$6856~\cite{Anumarlapudi2025MN}, DART/ASKAP\,J1832$-$0911~\cite{Wang2025Nat}, J1912$-$4410~\cite{Pelisoli2023NatAs}, and AR\,Sco~\cite{Marsh2016Nat}.
Some LPTs display distinctive emission characteristics: for example, ILT/CHIME\,J1634+44 shows transitions between nearly 100\% linear and 100\% circular polarization~\cite{Bloot2025AA}; CHIME\,J0630+25 exhibits a timing glitch of $\Delta F / F \sim 1.3\times10^{-6}$~\cite{Dong2025ApJLb}, where $\Delta F$ is the spin frequency change, and $F$ is the spin frequency; ASKAP\,J1935+2148 presents interpulse emission~\cite{Caleb2024NatAs}, i.e. a radio pulse separated from the main pulse by about 180 degrees in spin phase; and DART/ASKAP\,J1832-0911 displays pulsed X-ray emission that is phase-aligned with its radio pulses~\cite{Wang2025Nat}.

Although some LPTs are likely WD–M dwarf binaries or highly magnetized neutron stars, the physical nature of the remaining LPTs remains uncertain. Several theoretical models have been proposed to explain their radio emission: (1) WD–M dwarf binary systems, in which the emission can be produced by unipolar inductor–type magnetic interactions~\cite{Yang2025arXiv, Qu2025ApJ, Zhong2025arXiv} or by the interaction between the WD magnetosphere and the companion’s stellar wind~\cite{Horvath2025arXiv}; (2) Long-period magnetars~\cite{Beniamini2023MNRAS}, which require a revision of the classical pulsar death line, potentially through mechanisms such as twisted magnetic fields~\cite{Tong2023RAA, Cooper2024MNRAS} or fallback disks~\cite{Ronchi2022ApJ}; (3) Highly magnetized WD pulsars, where the emission is powered by WD spin-down~\cite{Katz2022ApSS}. Additionally, electron cyclotron maser emission (ECME) has been widely discussed as a potential driving mechanism of LPTs~\cite{Qu2025ApJ, Yang2025arXiv, Ferrario2025arXiv, Zhong2025arXiv}.

GPM\,1839$-$10 is an LPT with a period of about 21 minutes and has remained active for at least three decades~\cite{Hurley-Walker2023Nat}. Based on long-term archival data, its period derivative is constrained to $\dot{P}\lesssim3.6\times10^{-13}\,\mathrm{s\,s^{-1}}$, placing GPM\,1839-10 near the extreme edge of even the most generous pulsar death line predicted by classical emission models. 
The source has a dispersion measure (DM) of $\sim273.5\,\mathrm{pc\,cm^{-3}}$ and a rotation measure (RM) of $\sim531\,\mathrm{rad\,m^{-2}}$. The source exhibits complex pulse profiles and polarization behavior, including rapid variations in linear and circular polarization fractions and the presence of orthogonal polarization modes (OPMs)~\cite{Men2025SciA}, which have been widely observed in radio pulsars \cite{Philippov2022ARAA} and one FRB \cite{Jiang2024NSRev, Niu2024ApJ}.
Additionally, it shows intriguing down-drifting substructures in the dynamic spectrum reminiscent of those seen in repeating fast radio bursts (FRBs), as well as evidence for linear-to-circular polarization conversion and down-drifting polarization conversion~\cite{Men2025SciA}. 
Recent studies suggest that GPM\,1839$-$10 may be in a binary system with a possible companion, given the detection of an orbital period of 8.7\,hours~\cite{Horvath2025arXiv} and a possible M-type optical counterpart~\cite{Hurley-Walker2023Nat}, but no optical counterpart is confirmed~\cite{Pelisoli2025MNRAS}.

Cyclotron absorption occurs when electromagnetic waves propagate through a magnetized plasma and satisfy the cyclotron resonance condition~\cite{Luo2001MNRAS}. This phenomenon is typically observed in the X-ray spectra of highly magnetized neutron stars~\cite{Staubert2019AA}. However, in the radio band, evidence for such an effect has been reported only once, i.e. in MeerKAT observations of GPM\,1839$-$10~\cite{Men2025SciA}, making it a unique case among known radio sources. To further investigate the emission mechanism and physical nature of GPM\,1839$-$10, we conducted observations using the Five-hundred-meter Aperture Spherical Telescope (FAST). The details of the observation and data analysis are presented in \SEC{observation}, the results are described in \SEC{results}, and the discussion is provided in \SEC{discussion}. The conclusions are summarized in \SEC{conclusion}.

\section{Observation and data processing}\label{sec:observation}
The observations were carried out with FAST using the 19-beam L-band receiver, which covers a frequency range from 1.0 to 1.5\,GHz~\cite{Jiang2019SCPMA, Jiang2020RAA}, under the open-time proposal (project ID: PT2024\_0124). Only the data from the central beam were recorded in search mode, stored in the PSRFITS format with dual polarization. A total of four hours of observations were divided into five sessions, as summarized in \TAB{observation}. At the beginning of each session, a three-minute observation with noise injection was performed for polarization calibration.
\begin{table}[H]
\centering
\footnotesize
\begin{threeparttable}\caption{FAST observation of GPM\,1839-10}\label{tab:observation}
\doublerulesep 0.1pt \tabcolsep 13pt 
\begin{tabular}{cccc}
\toprule
  Start time & Length & Sample Time & Channel\\
  (UTC) & (s) & ($\mu$s) & (number) \\\hline
  2024-10-07 10:35:00 & 2100 & 98.304 & 8192\\
  2024-11-15 08:01:00 & 2100 & 98.304 & 8192\\
  2024-12-04 06:50:00 & 2100 & 98.304 & 8192\\
  2024-12-10 05:16:00 & 2100 & 98.304 & 8192\\
  2025-01-17 02:50:00 & 3000 & 98.304 & 8192\\
\bottomrule
\end{tabular}\end{threeparttable}
\end{table}

\begin{figure*}
\centering
\includegraphics[width=\textwidth]{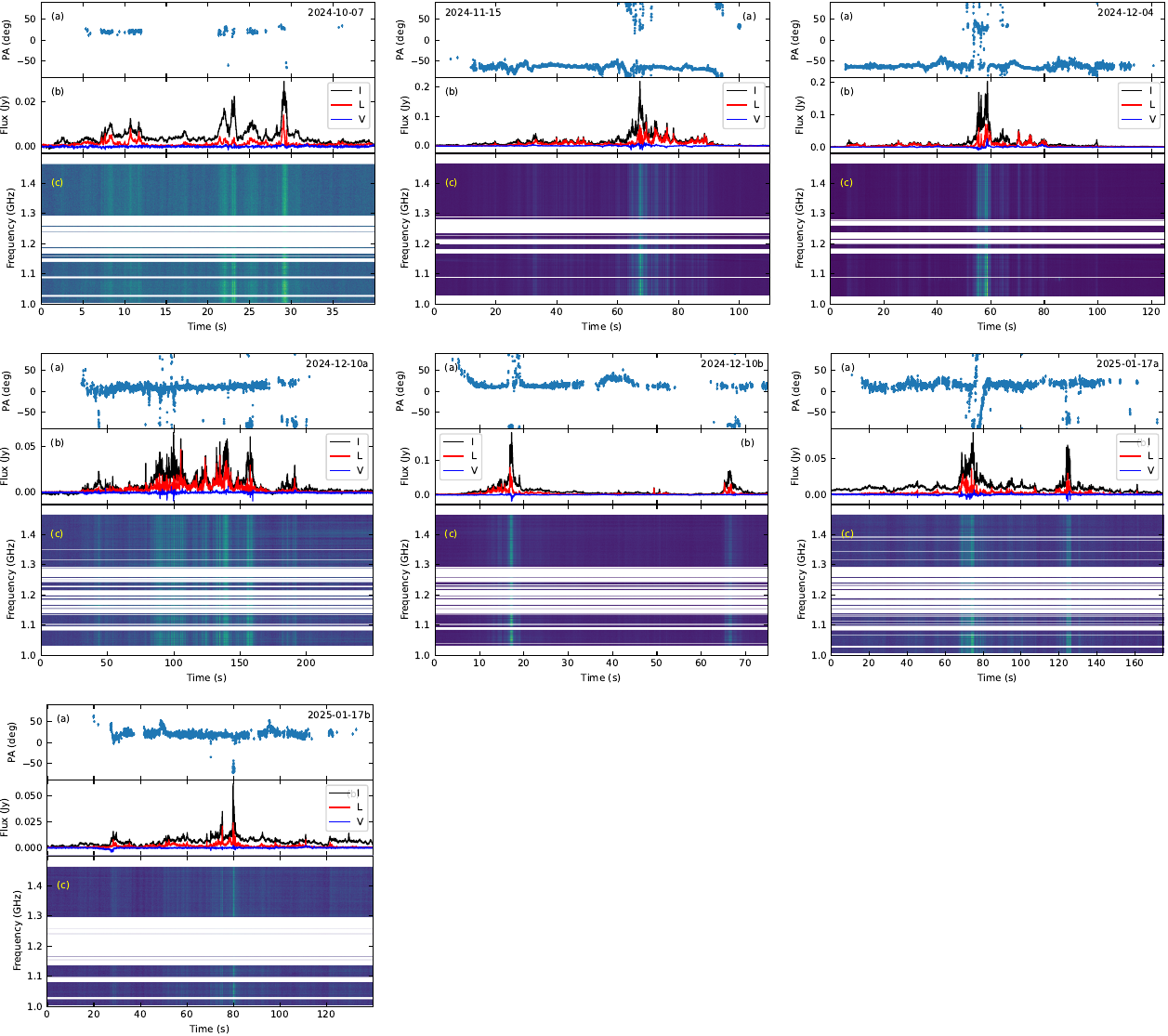}
\caption{Radio pulses from GPM\,1839$-$10 detected with FAST. (a) PA variation, (b) total intensity (black), linearly polarized intensity (red), and circularly polarized intensity (blue), and (c) dynamic spectra of the radio pulses are shown.} 
\label{fig:radio_pulses}
\end{figure*}

The data were first processed using {\sc DSPSR}~\cite{Straten2011PASA}, which generated archive files containing time-scrunched dynamic spectra for visual inspection and identification of radio pulses in the observation data. Data segments containing detected pulses were then extracted with {\sc replot\_fil} in {\sc TransientX}~\cite{Men2024AA}, producing corresponding archive files for the selected time ranges. Narrow-band radio frequency interference (RFI) was mitigated using {\sc clean\_archive.py} in {\sc PulsarX}, which removes narrowband RFI signals based on statistical analysis of each frequency channel~\cite{Men2023AA}. Because the gain and phase responses can differ between the two orthogonal polarization signal paths, polarization calibration was required. The observations with injected calibration noise were used to measure the Mueller matrix response for each frequency channel, enabling polarization calibration. To perform the calibration, we folded the noise diode data using the modulation period and manually removed RFI-contaminated channels with {\sc pazi} in {\sc PSRCHIVE}~\cite{Straten2012ART}. The GPM\,1839$–$10 data were then calibrated using the cleaned noise diode observations with {\sc pac} in {\sc PSRCHIVE}. We subsequently performed RM synthesis~\cite{Brentjens2005AA} to determine the RM of each radio pulse and corrected for Faraday rotation using the measured RMs. Additionally, {\sc IONFR}~\cite{Sotomayor2013AA} was applied to mitigate the ionospheric contribution to the RM. The DM was estimated with {\sc DM\_phase}~\cite{Seymour2019ascl}, and the data were finally dedispersed using the measured DMs.

\begin{figure}[H]
\centering
\includegraphics[scale=0.8]{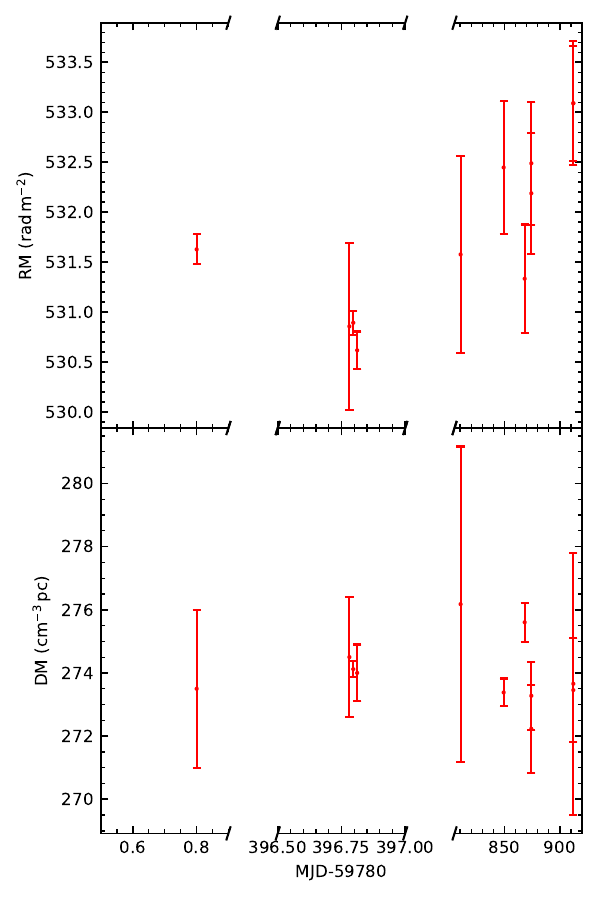}
\caption{RMs and DMs as measured at different epochs. The values are measured using the observations in this work and previous work~\cite{Hurley-Walker2023Nat, Men2025SciA}.} 
\label{fig:rm_dm_change}
\end{figure}

\section{Results}\label{sec:results}

\begin{figure*}
\centering
\includegraphics[width=\textwidth]{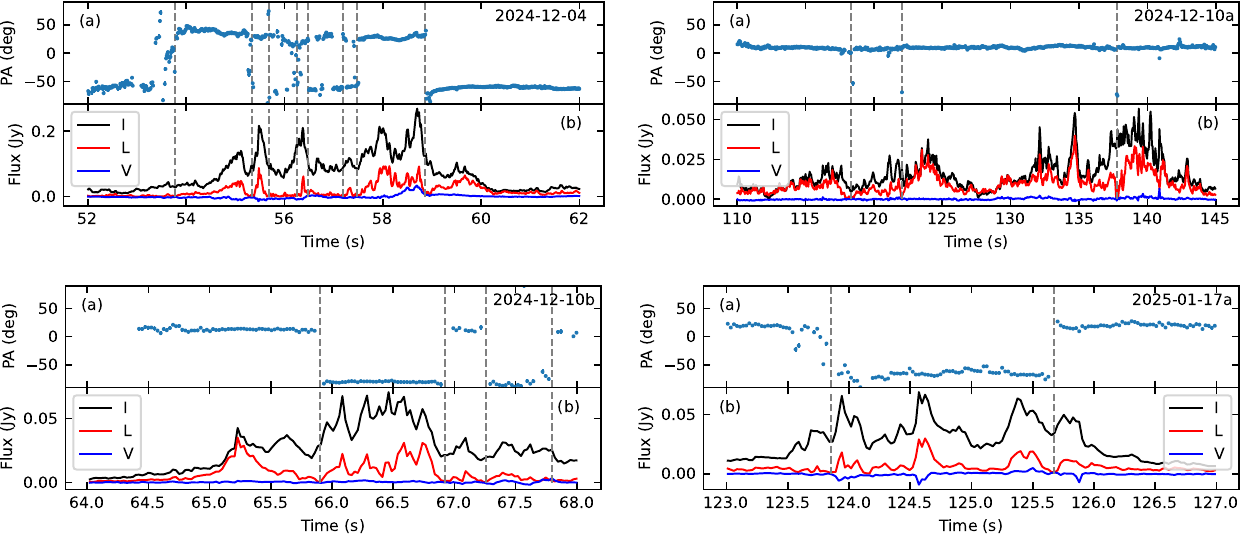}
\caption{Correlation between OPM switches and decreases in polarized intensity. (a) PA variation, (b) total intensity (black), linearly polarized intensity (red), and circularly polarized intensity (blue) are shown. The gray dashed lines mark the times when OPM switches coincide with decreased polarized intensity.
} 
\label{fig:OPM}
\end{figure*}

A total of seven rotational periods exhibited clear radio emission, as shown in \FIG{radio_pulses}. The radio profiles display complex structures that vary between epochs, with durations of tens of seconds, consistent with previous observations with MeerKAT~\cite{Men2025SciA}. The linear polarization fraction occasionally reaches nearly 100\%, and frequent OPM switches in the polarization angle (PA) are observed, although the PAs remains mostly flat across the pulses. Furthermore, the measured RMs of the radio pulses in these observations show a slight increase but remain statistically insignificant compared to previous observations, and the DMs also show no significant variation, as shown in \FIG{rm_dm_change}. Additionally, the polarization fraction approaches zero during these OPM switches, as illustrated in \FIG{OPM}, suggesting that the switches are caused by the incoherent superposition of the two OPMs. A detailed discussion of this phenomenon is provided in \SEC{discussion}. 

\begin{figure*}
\centering
\includegraphics[width=0.8\textwidth]{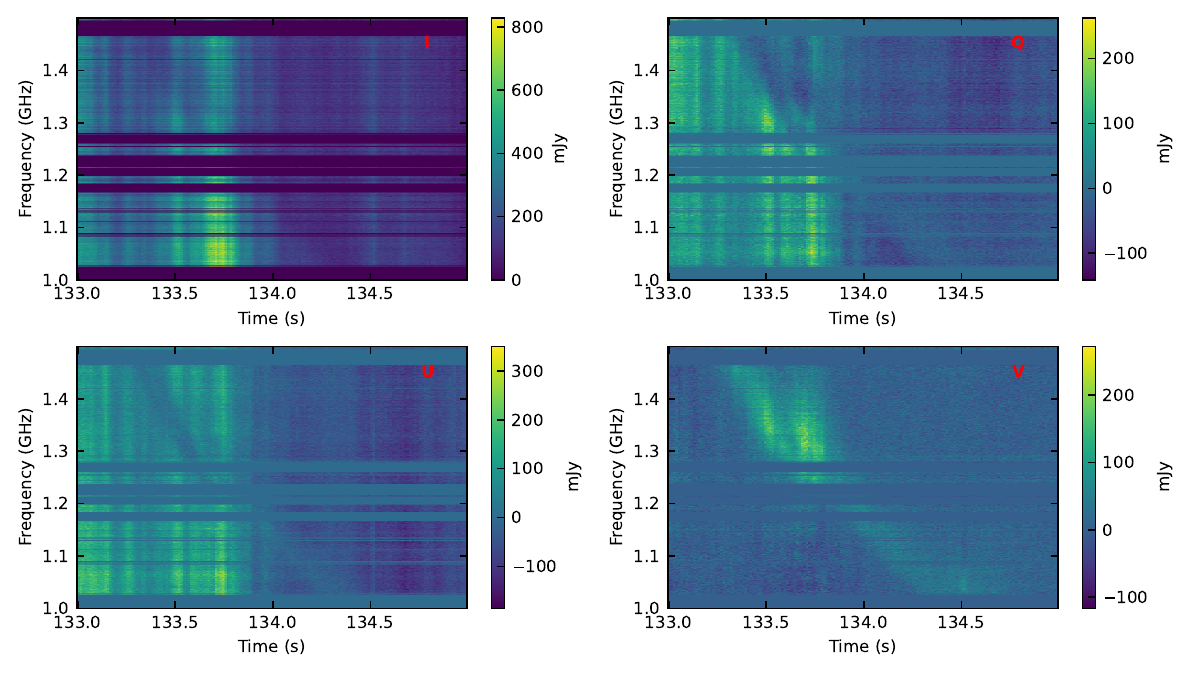}
\caption{Faraday de-rotated dynamic spectra for the Stokes parameters (I, Q, U, and V) showing signatures of cyclotron absorption. The data are from the FAST observation conducted on 2024 December 4. Frequency-dependent decreases in total intensity (I) and linear polarization intensity (Q and U), accompanied by an increase in circular polarization (V), are detected.
} 
\label{fig:down_drifting}
\end{figure*}

\begin{figure*}
\centering
\includegraphics[width=\textwidth]{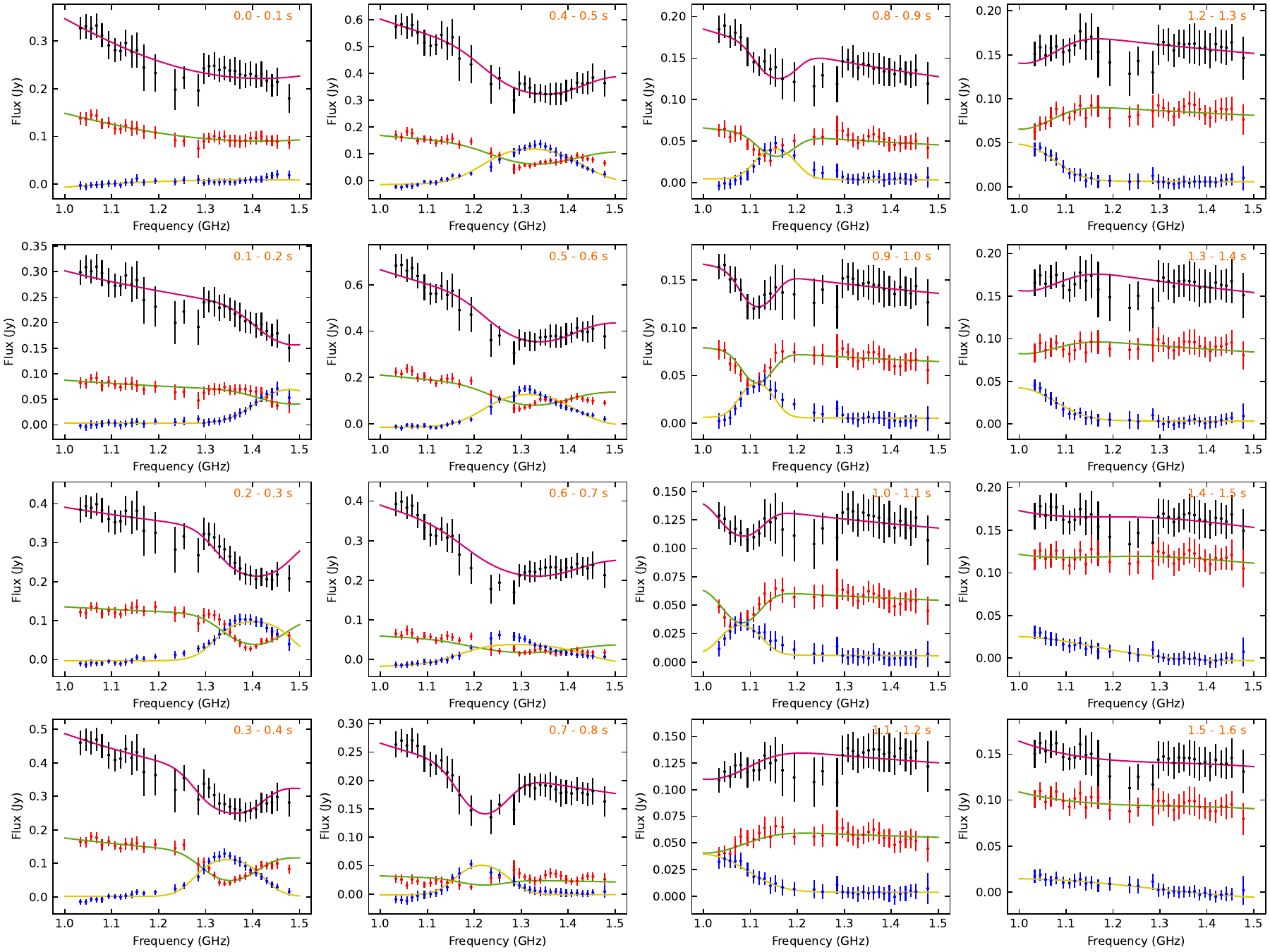}
\caption{Results of fitting the variation of total intensity (black dots), linearly polarized intensity (red dots), and circularly polarized intensity (blue dots) as a function of radio frequency over multiple 100\,ms time intervals. The time range for each panel is indicated at the top right, with the start time corresponding to 133.2\,s in \FIG{down_drifting}.} 
\label{fig:down_drifting_fitting}
\end{figure*}

In the observation conducted on December 4th, 2024, we detect a frequency-dependent polarization variation within a down-drifting frequency band, where both the total intensity and linear polarization intensity decrease, while the circular polarization intensity increase, as shown in \FIG{down_drifting}. This emission feature resembles the phenomenon previously reported in MeerKAT observations~\cite{Men2025SciA}, but appears more pronounced in the present data. 
These characteristics are qualitatively consistent with expectations from cyclotron absorption in a magnetized plasma.
In a strong background magnetic field, resonant cyclotron absorption occurs when the radio wave angular frequency satisfies $\omega/{\cal D}=\omega_B$, where the electron cyclotron frequency is $\omega_B=eB/m_ec$ and ${\cal D}=1/\gamma(1-\beta\cos\theta)$ is the Doppler factor. 
Here $\theta$ is the angle between the wave’s momentum and the electron velocity, $\beta=v/c$ is the normalized speed of the electron, and $\omega$ is angular frequency of radio wave.
Since left- and right-hand circularly polarized waves interact differently with magnetized charges, cyclotron absorption naturally introduces frequency-dependent and polarization-dependent attenuation, preferentially reducing the linearly polarized intensities while enhancing net circular polarization~\cite{Luo2001MNRAS, Qu2023MNRAS} (See \SEC{discussion} for detailed discussions).

To characterize this behavior, we model the polarization-dependent absorption by decomposing the incident radiation into two orthogonal circularly polarized R- and L-mode. The mode optical depths are $\tau_{\rm R,L}(f)=\tau_{\rm R/L,0}\exp[-(f-f_c)^2/w_f^2]$ and $\Bar{\tau}=(\tau_R+\tau_L)/2$.
The Stokes $I$ and $V$ are 
\begin{equation}
I(f)=\left(\frac{f}{f_0}\right)^{-\alpha}\left[I_\mathrm{R} e^{-\tau_R(f)}+I_\mathrm{L} e^{-\tau_L(f)} + I_{\mathrm{un}}\right],
\end{equation}
\begin{equation}
V(f)=\left(\frac{f}{f_0}\right)^{-\alpha}\left[I_\mathrm{R} e^{-\tau_R(f)}-I_\mathrm{L} e^{-\tau_L(f)}\right],
\end{equation}
Since
\begin{equation}
Q(f)=Q_0\left(\frac{f}{f_0}\right)^{-\alpha}e^{-\bar\tau(f)}, \ 
U(f)=U_0\left(\frac{f}{f_0}\right)^{-\alpha}e^{-\bar\tau(f)}.
\end{equation}
We have
\begin{equation}
L(f)=\sqrt{Q(f)^2+U(f)^2}
= L_0\left(\frac{f}{f_0}\right)^{-\alpha}e^{-\bar\tau(f)}.
\end{equation}

\begin{figure}[H]
\centering
\includegraphics[scale=0.8]{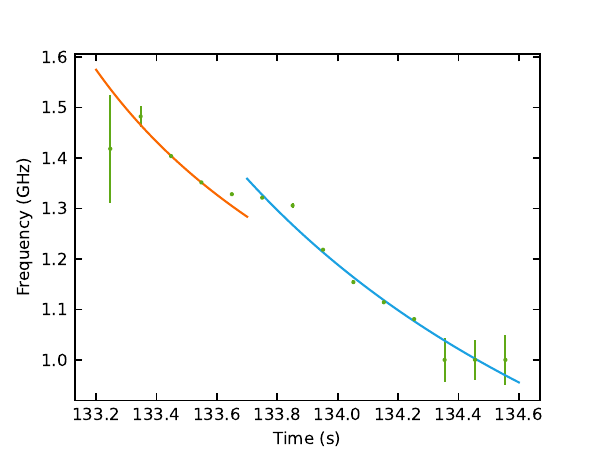}
\caption{Fitting of the drift rate of the cyclotron absorption feature. The fit is applied separately to two segments splitting at 133.7 ms, corresponding to the two drifting bands visible in \FIG{down_drifting}.} 
\label{fig:drift_relation_fit}
\end{figure}

\begin{figure}[H]
\centering
\includegraphics[scale=0.8]{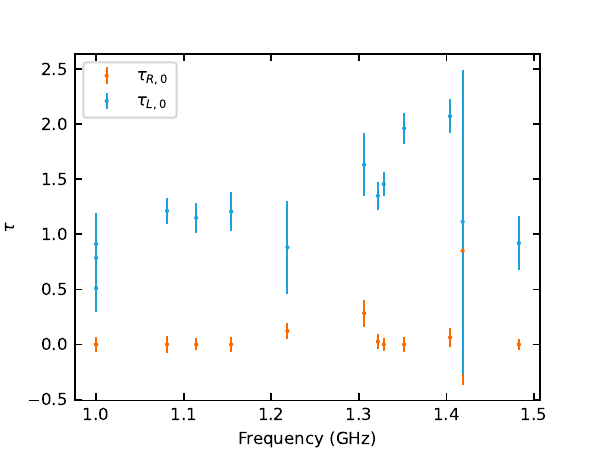}
\caption{Optical depth as a function of frequency for right- and left-handed circular polarization in the cyclotron absorption feature.} 
\label{fig:tau_relation_freq}
\end{figure}

$I_\mathrm{R}$, $I_\mathrm{L}$, and $L_0$ are the unabsorbed intensities for right-hand circular, left-hand circular, and linear polarization at the reference frequency $f_0 = 1,\mathrm{GHz}$. $\alpha$ is the spectral index of the intensity, and $I_\mathrm{un}$ is the unpolarized intensity. The slight drop in total intensity around 1.25\,GHz is likely due to the receiving system's response. The Gaussian-shaped intensity decrease as a function of frequency is characterized by its central frequency $f_c$ and width $w_f$. The fitting results are displayed in \FIG{down_drifting_fitting}. The results show that the total and linearly polarized intensities decrease, while the circularly polarized intensity increases within a narrow frequency band in each time interval. This band is well described by a Gaussian profile, and its central frequency decreases with time. Subsequently, we fitted the evolution of $f_c$ over time using the model,
\begin{equation}
    f = f_0 \left( \frac{t-t_0}{(D/241)\ \mathrm{s}}\right)^{-\frac{1}{\beta}}\,,\\
\end{equation}
where $D$ is a dimensionless parameter with a meaning analogous to the DM, and $\beta$ is the frequency-dependent relation index of the drifting. The polarization spectrum exhibits two down-drifting bands. As they separate around 133.7 ms (see \FIG{drift_relation_fit}), we fitted the $f_c$--$t$ relationship for each band separately. The fitting results, displayed in \FIG{drift_relation_fit}, yield drift rates $D$ of $706^{+145}_{-186}$ and $916^{+148}_{-582}$ with $\beta$ of $1.3^{+3.7}_{-0.5}$ and $0.7^{+0.9}_{-0.1}$, corresponding to drift rates of about 550\,MHz/s at 1.4\,GHz and 400\,MHz/s at 1.1\,GHz, respectively. Interestingly, the latter drifting rate aligns with results from a previous MeerKAT study conducted in the UHF band, which spanned from 544\,MHz to 1088\,MHz. Furthermore, only the left-hand circular polarization component shows significant absorption in \FIG{tau_relation_freq}, a result indicative of cyclotron absorption.

\section{Discussions}\label{sec:discussion}

The PA jumps occur frequently in GPM\,1839$-$10 and share similar characteristics with those observed in radio pulsars and FRBs. 
Figure~\ref{fig:OPM} suggests that two distinct polarization modes dominate before and after each PA jump. 
Theoretically, PA jumps can occur from either coherent or incoherent superposition of two waves. 
In the coherent superposition case, PA jumps occur when the linear polarization fraction reaches a minimum while the circular polarization peaks, provided that the two initial modes are linearly polarized and the total polarization degree is conserved~\cite{Qu2026ApJ}. 
On the other hand, incoherent superposition naturally leads to depolarization, which appears more consistent with the PA jumps detected in GPM\,1839$-$10. The observed PA jumps occur on a timescale of roughly 0.1 s, implying an emission region size of order $\sim 10^9\ \rm cm$ which is well inside the magnetosphere of the central engine.

The detected frequency-dependent attenuation of linear polarization intensity, accompanied by enhanced circular polarization, is suggestive of resonant cyclotron absorption.
This propagation effect occurs when the radio wave frequency satisfies the resonance condition $\omega_B'=\omega'$ in the comoving frame of the particle.
For GPM\,1839$-$10, we divide the burst into sixteen 0.1-s time bins in Figure~\ref{fig:down_drifting_fitting}.
Cyclotron absorption is evident at different time intervals, occurring when the circular polarization increases (blue points) and the total intensity decreases (black points).
The inferred resonant frequencies span nearly the entire observing band. 
The local magnetic field strength can be estimated from the cyclotron resonance condition
\begin{equation}
B_{\rm bg}=\frac{2\pi m_ec}{e}\gamma(1-\beta\cos\theta)\nu.
\end{equation}
For $\gamma=5$, this gives $B_{\rm bg}\simeq (36 \ {\rm G}) \ \nu_9$ for aligned propagation ($\theta=0$), $\nu_9$ is the radio frequency in the unit of GHz. 
While for a representative oblique case $\theta=30^\circ$ the inferred field is larger by a factor of $\sim7$, i.e. $B_{\rm bg}\sim (10^2 \ {\rm G}) \ \nu_9$.
We note that for mildly relativistic electrons, modest variations in $\theta$ introduce order unity corrections to the Doppler factor.
Here $\gamma$ can be regarded as a characteristic Lorentz factor of the resonant particles.
In a realistic magnetospheric plasma, the particle energy distribution could be broad, and cyclotron absorption at each frequency is dominated by a subset of electrons that locally satisfy the resonance condition.

It should be pointed out that the cyclotron absorption frequency decreases within seconds timescales, which indicates a typical light crossing time region with $\lesssim 10^{10} \ \rm cm$, which is smaller than the separation distance of the binary system, i.e. $a\sim1.5\times10^{11}\,\mathrm{cm} \left(\frac{P_b}{8.7\,\mathrm{hr}}\right)^{2/3}\left(\frac{M}{1\,M_\odot}\right)^{1/3}$, if the GPM\,1839-10 is a binary origin. This separation is smaller than the light cylinder radius, i.e. $r_l=\frac{P\,c}{2\pi}\sim6\times10^{12}\,\mathrm{cm}$. In the background dipole field configuration, $\theta$ will gradually increase. Thus, the downward drifting of the resonant cyclotron absorption frequency indicates that the wavefront progressively encounters regions of decreasing magnetic field strength.

At least four known LPTs are associated with M-dwarf companions~\cite{Ruiter2025NatAs, Hurley-Walker2024ApJL, Pelisoli2023NatAs, Marsh2016Nat}. GPM\,1839-10 may similarly reside in such a binary system based on previous studies ~\cite{Hurley-Walker2023Nat, Horvath2025arXiv}. In this scenario, the absorption could occur in the vicinity of a weakly magnetized companion, accounting for both the magnetic field strength required by the observed cyclotron absorption feature. However, the orbital phase corresponding to the absorption feature is about 0.85, at which the companion would not be in the observer’s line of sight. This inconsistency needs to be examined in detail with additional observations. An alternative scenario posits that the cyclotron absorption occurs within the magnetosphere of a central compact star. Under this hypothesis, the distance from the compact star, estimated from the measured magnetic field strength assuming a dipolar magnetic field model, is $\sim(1.4\times10^{11}\,\mathrm{cm}) \nu_{9}^{-1/3}B_{8}^{1/3} r_{9}$, where $B_8$ the magnetic field strength in units of $10^8\,\mathrm{G}$ and $r_9$ is the radius in units of $10^9\,\mathrm{cm}$. Such magnetic field strengths can be achieved within the magnetospheres of either neutron stars or magnetized white dwarfs. Additionally, the RMs measured in these observations are consistent with previous MeerKAT measurements~\cite{Men2025SciA}, indicating that the magneto-ionic environment is stable. This stability suggests that the system is unlikely to host a highly magnetized companion with a dynamically varying magneto-ionic environment, such as those found in spider pulsars~\cite{Wang2023ApJ} or PSR 1259-63, where the RMs are known to show large variations~\cite{Johnston1992ApJ, Wang2023ApJ}.

\section{Conclusions}\label{sec:conclusion}
In this work, we present observations of GPM\,1839-10 conducted with the FAST telescope in the L-band, detecting a total of seven radio pulses across seven distinct periods. The radio pulses exhibit complex profiles but are characterized by a flat position angle (PA) variation. Additionally, the OPM switched frequently across the pulses, which was consistently associated with decreases in the polarized intensity. This behavior can be well explained by the incoherent summation of OPMs. Moreover, we detected clear evidence of cyclotron absorption in one radio pulse, manifested by a decrease in the total and linear polarized intensity, accompanied by an increase in the circular polarized intensity. This feature appeared in a down-drifting sub-band, with a drift rate consistent with previous MeerKAT observations in the UHF band. The cyclotron absorption feature could originate in a magneto-ionic region with a lower limit magnetic field strength of tens of Gauss, which can be close to a weakly magnetized companion. Furthermore, the RMs are consistent with previous MeerKAT observations, indicating a stable magneto-ionic environment. Such stability suggests that, in the case of a binary system, the companion is likely weakly magnetized. Future observations will be crucial to further test this scenario.

\Acknowledgements{This work is supported by the Max Planck society. This work made use of data from FAST—a Chinese national megascience facility, built and operated by the National Astronomical Observatories, Chinese Academy of Sciences. We thank the anonymous referees for very helpful comments. Finally, R.L. wishes to posthumously acknowledge his late father, Luo Kongdao, for the unwavering support, belief, and lifelong love.}

\InterestConflict{The authors declare that they have no conflict of interest.}

\end{multicols}
\end{document}